# DATA WORK IN EGYPT

# WHO ARE THE WORKERS BEHIND ARTIFICIAL INTELLIGENCE?

DIPLAB REPORT 2025-2026

# HOW TO CITE THIS REPORT

Myriam Raymond, Lucy Neveux, Antonio A. Casilli, Paola Tubaro (2025). *Data Work in Egypt. Who are the Workers Behind Artificial Intelligence*. DiPLab Report.

# ACKNOWLEDGMENTS

This report was made possible thanks to the intellectual guidance, methodological support, and sustained encouragement of the DiPLab research collective, whose work on digital labor ecosystems continues to shape our understanding of platform-mediated work.

I am particularly grateful for the methodological directions and analytical frameworks developed within DiPLab, which informed both the design of this study and the interpretation of its findings.

I also acknowledge the financial support that contributed to the realization of this work, including the funding streams that enabled field preparation, data collection, and collaborative analysis.

My sincere thanks go to my research assistant, Lucy Neveux, for her rigorous handling of the dataset, careful data analysis, and continuous work in structuring and organizing the empirical material. Her contribution was essential to the completion of this report.

*Myriam Raymond*

DiPLab

# KEY FINDINGS

**76%** of data workers identify as **men** (*n=484/636*)

**74%** of data workers are **between 18 and 34 years old** (*n=468/636*)

**3 out of 4** data workers **depend on platforms to earn enough** to pay their bills (*n=514/683*)

The **average monthly** salary from platforms is EGP 2793,37 **(USD 58,76)**. Minimum wage in the country is EGP 7000 (USD147) per month (*N=680*)

On average, data workers spend **12.3 hours** on platforms **per week**

**15%** of respondents work **7 days a week** (*n=106/709*)

**60%** of data workers have a **Bachelor of Science or a technical degree** (*n=409/671*)

**83%** of data workers reported that their **primary motivation** for working on platforms is **financial necessity** (*n=557/668*)

DiPLab

# WHAT IS DATA WORK ?

**Data work** refers to a form of digital labor facilitated by online platforms, where workers perform repetitive, on-demand tasks that are standardized into units of service and compensated on a piecemeal, hourly, or project basis (Graham et al., 2017; Posada and Miceli, 2022). According to a World Bank estimate, this type of "online gig work" provides income opportunities for millions of workers but also exposes them to the risk of economic marginalization (Datta et al., 2023).

These are known as **microtasks.** They represent the breakdown of complex digital processes into small, paid assignments distributed to a dispersed workforce (Gray & Suri, 2019; Casilli, 2025). Microtasks are often paid only a few cents in local currency or U.S. dollars. They represent a highly fragmented and informal type of work that typically lacks labor and social protections.

Following the **release of ChatGPT in late 2022** and the subsequent wave of news stories about the "hidden cost of AI," data work has become a central topic in public debate. Whether referred to as "moderators", "data annotators", or "AI reviewers", they all belong to a burgeoning category of workers who shape contemporary modes of information production. Despite its hidden and precarious nature, their activity is a critical component of the digital economy, providing the **human input necessary to train, test, and refine Artificial Intelligence models** (Tubaro et al., 2020).

These activities highlight a rapidly evolving technological paradigm where, despite predictions of the "end of work," human labor remains essential to the functioning and optimization of algorithmic systems.

DiPLab

# AI'S UNDERPAID DATA WORKFORCE

With AI becoming more prevalent, public opinion mostly revolves around powerful machine learning models and their designers, software engineers, and data scientists. However, these well-known and well-paid professionals represent only a small fraction of AI's global workforce. The **majority of the labor required to produce AI** is carried out by **underpaid and precarious data workers on platforms,** performing microtasks from corporate facilities, informal structures, and their homes.

All machine learning relies on **generating, classifying, preparing, verifying, and annotating data**. And this is where human work is increasingly needed. Training and data annotation work is **typically outsourced by tech companies in the Northern hemisphere to data workers in the Majority World**. While some are recruited by companies, call centers, or startups, most workers access global microtasking platforms like Microworkers, UHRS, Appen, and ClickWorker, operating as independent "work providers."

By focusing on the latter category, in this report we aim to understand Egypt's place within the global supply chain of all AI development.

DiPLab

# WHAT PLATFORMS ARE USED BY EGYPTIAN DATA WORKERS? (1/2)

Arguably the first online platform entirely devoted to data work, Amazon Mechanical Turk was launched in 2005. Since then **data work platforms** have proliferated globally, creating new forms of online labor. In previous DiPLab studies focused on France (2019), Brazil (2023), and 9 European countries (2024) we identified the existence of several platforms, each with different objectives that range from data training for machine learning to remote usability tests, and even the creation of fake profiles to boost social media (like in click farms).

In Egypt alone, our research identified **11 major platforms**—including Microworkers, Contra, Telus, OneForma, Toloka, Fiverr, Clickworker, Upwork, Appen, UHRS, and Remotasks—each serving different purposes, from **training data for machine learning** and **content moderation** to **remote usability testing** and even **social media manipulation** (e.g. the creation of fake profiles and engagement in so-called click farms). Survey respondents also reported additional platforms—such as SproutGigs, Outlier AI, Mostaql, uTest, UNU, Timebucks, Pawns, CortalyCash, ySense, Honeygain, and Populii.[1]

The **range of microtasks** performed is increasingly diverse. Workers now engage in:
- **Data-centric tasks**, such as image and text annotation, audio and video transcription, ad evaluation, and content labeling for AI training.
- **Testing and evaluative tasks**, including app and website testing, product reviews, game testing, and market research surveys.
- **Engagement and promotional tasks**, such as viewing videos, subscribing to channels, liking or commenting on posts, and generating synthetic engagement across social media platforms like YouTube, Facebook, Instagram, TikTok, and Spotify.

[1] Data of the survey may reflect a platform-specific bias regarding familiarity with specific digital platforms , given that the survey was conducted on Microworkers and Clickworker.

DiPLab

# WHAT PLATFORMS ARE USED BY EGYPTIAN DATA WORKERS? (2/2)

To different degrees, all of these tasks qualify as data work. They generate, structure, or evaluate the data that STARA systems (Smart Technologies, AI, Robotics, and Algorithms) rely on to function and improve. These tasks transform unstructured content into structured datasets while also refining and optimizing solutions. Even engagement actions such as views, likes, and comments create behavioral datasets that directly train recommendation, ranking, and ad-targeting algorithms.

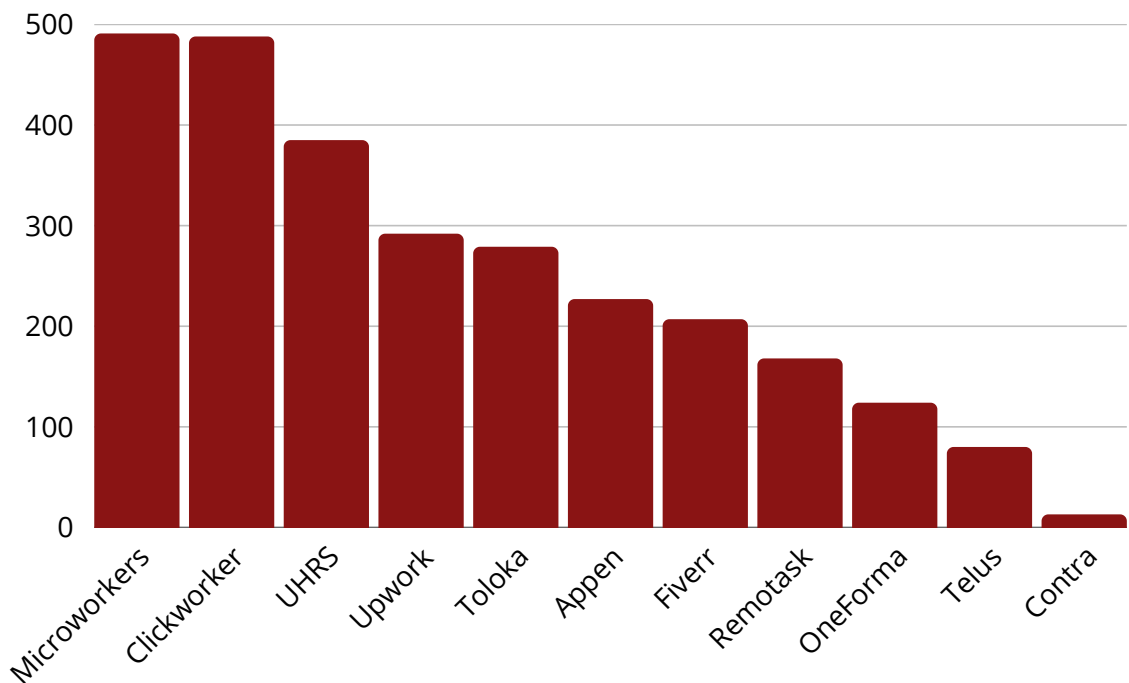

Figure 1 - Number of respondents having worked on each platform

The most used platform is UHRS[1] (385 respondents affirmed having previously worked on this platform). Upwork is also quite popular, and so is Microworkers. In most cases, the respondents had been working on these platforms for less than 7 months and they had found out about them through internet search.

---

[1] The Universal Human Relevance System (UHRS) is a Microsoft crowdsourcing platform that offers small, quick micro-tasks ("HitApps") to human workers (called "judges") to help improve AI and machine learning models. Workers access UHRS through third-party vendors like Clickworker, Appen, or OneForma, which manage their registration, payments, and support. By completing these micro-tasks, "judges" provide valuable human input that helps AI systems better understand context, recognize objects, and assess relevance — ultimately improving their accuracy and performance.

DiPLab

# WHO ARE THE EGYPTIAN DATA WORKERS? (1/3)

Egyptian data workers are mostly young, between **18 and 24 years old** (**34%**, n=216/635) and between **25 and 34 years old (40%**, n=252/635). They mostly identify as **men** (**76%**, n=484/636) and **single** (**59%**, n=359/610).

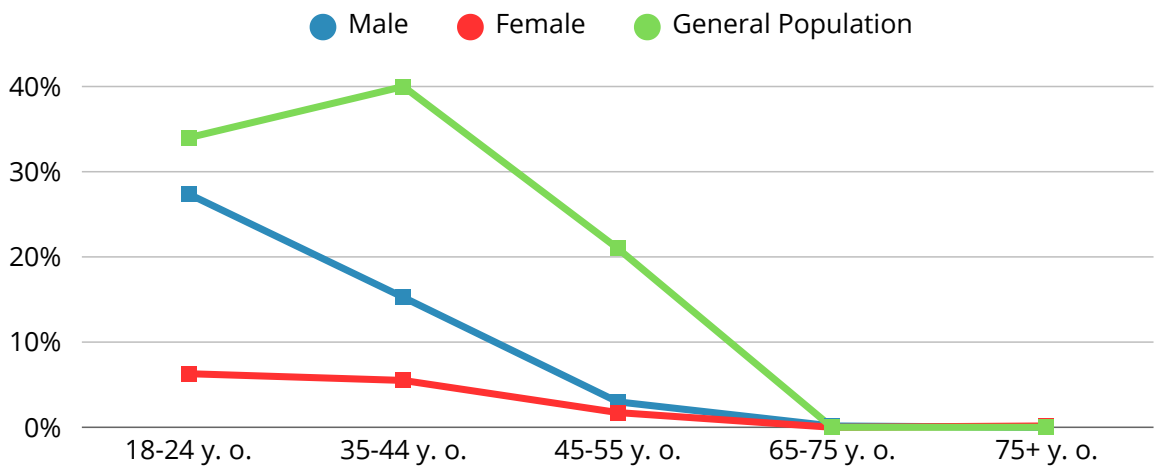

Figure 2 – Gender of data workers by age bracket compared to Egypt's general population

They mostly live in **Cairo and Alexandria**, but the sample is well distributed across regions, as respondents come from all over Egypt. This reflects broader **urban patterns observed internationally** (Casilli 2019; Braesemann et al. 2022). While data work can theoretically be performed from anywhere, in practice it remains tied to population and income distribution. Urban areas, particularly large cities, have a higher proportion of working-age individuals and better internet infrastructure, that favor the concentration of data workers. In contrast, lower income rural regions often have fewer residents in prime working age and limited connectivity, which restrict access to these opportunities.

# WHO ARE THE EGYPTIAN DATA WORKERS? (2/3)

Online platforms often present tasks that may seem **deceptively simple**, such as paraphrasing a sentence, transcribing a conversation, or providing words to describe an image. However, the Egyptian workers recruited to perform those tasks are **highly educated individuals**.

Additionally, workers acquire new skills online to complete tasks more quickly and effectively. This is coherent with our finding that **most workers have a bachelor of science or another technical degree** (60%, n=409/671). Generally, spouses also tend to have a scientific background; 66% of spouses whose education level was specified have a bachelor of science or a technical degree (n=171/261).

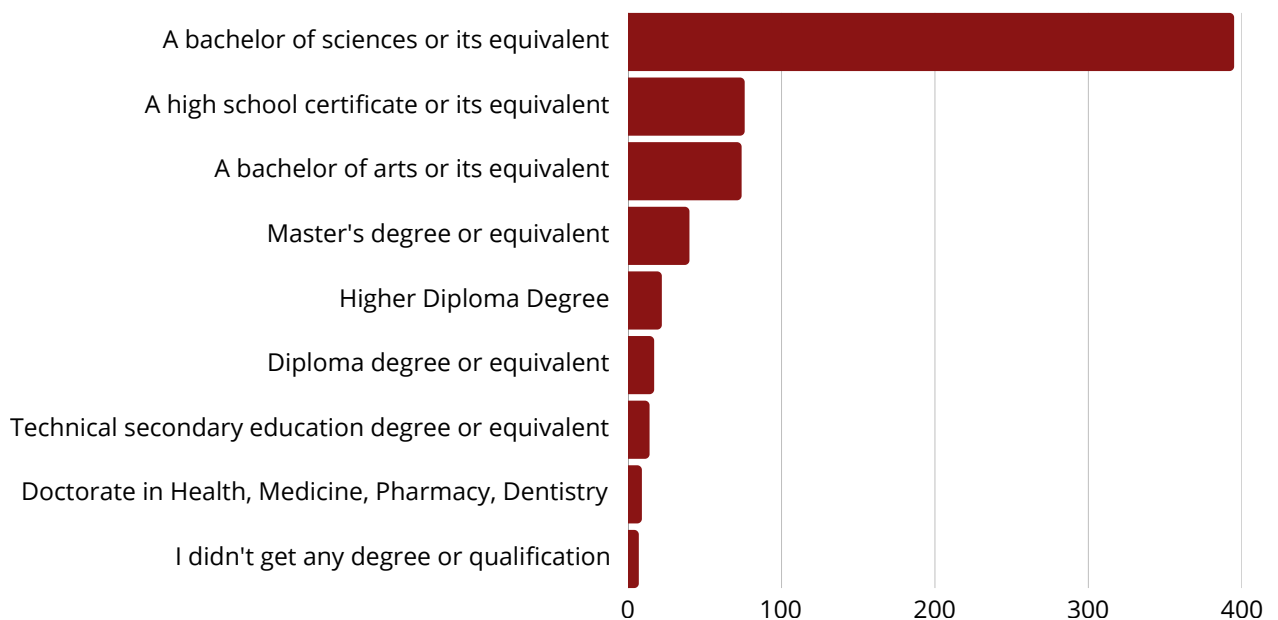

Figure 3 - Educational level of Egyptian data workers

# WHO ARE THE EGYPTIAN DATA WORKERS? (3/3)

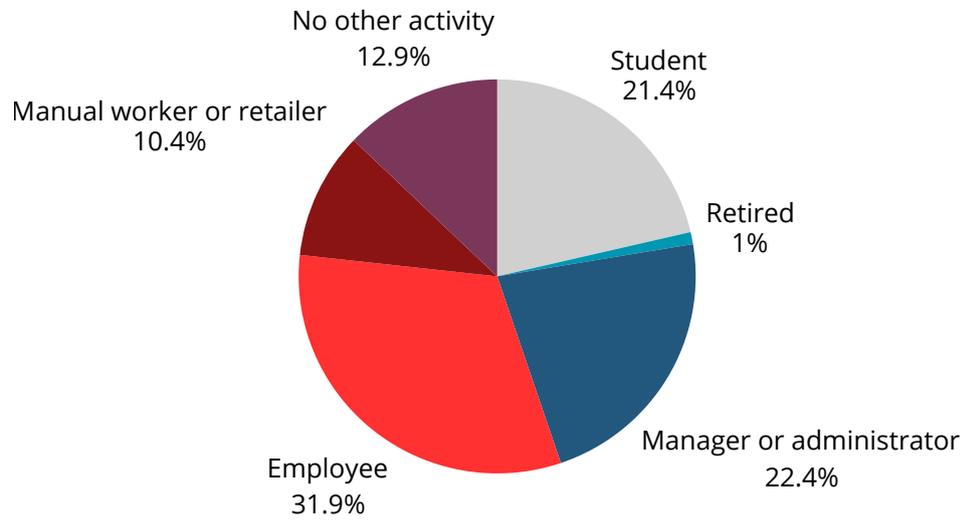

Figure 4 – Main activity of Egyptian data workers

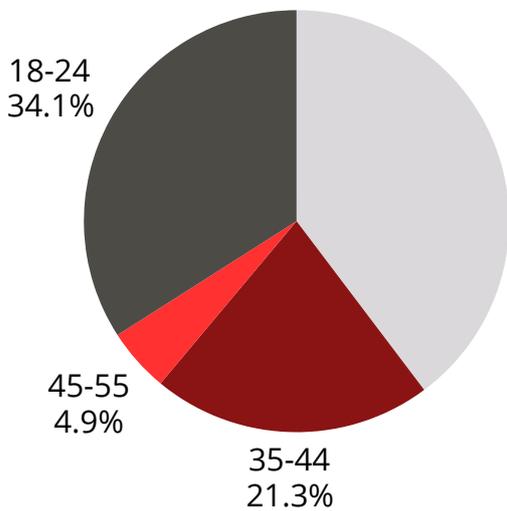

Figure 5 – Age distribution

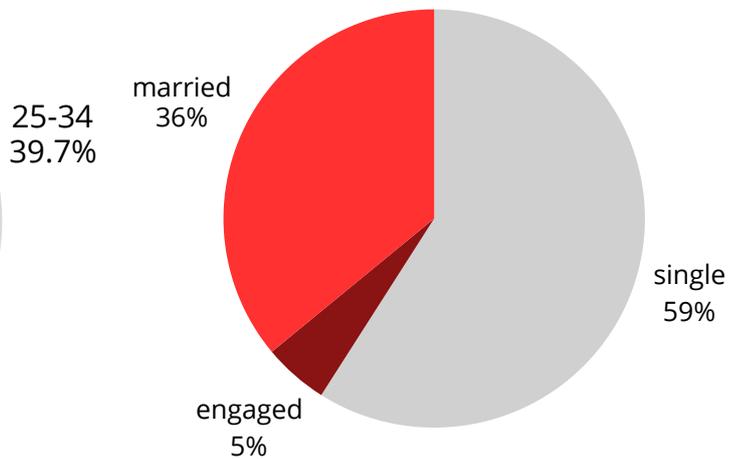

Figure 6 - Marital status

# HOW MUCH DO EGYPTIAN DATA WORKERS EARN? (1/3)

- In addition to their data work, nearly half of the respondents **have another job**. Of these, **almost two-thirds hold full-time positions**.

- The **average monthly income on data work platforms is EGP 2,793.37 (USD 58.76).** For reference, minimum wage in Egypt is EGP 7,000 per month (= USD 149) and the average per month salary as of 2025 is roughly EGP 9,200 (= USD 195).

- An **Egyptian data worker gains on average USD 1.22 per hour**, which is comparable to the average Egypt's hourly rate of USD 1.01[1].

- Considering all sources of income (data work, wages from other jobs, pensions etc.), **47%** (n=306/658) **of data workers' households earn less than EGP 8,000 per month** (= USD 168.56).

- Across all platforms, data workers are active about equally on all days except Sunday, which sees less activity. This seems to indicate that they **mostly work for companies and clients primarily located in Europe, the U. S. and Sunday-observant Asian Countries such as China or Japan.**

- On average, workers spend **106 minutes on platforms per day**.

- 120 respondents reported **using unofficial intermediaries based in the UAE to transfer funds** in order to receive the outstanding balance owed to them from the platforms in Egypt.

---

[1] Calculated as average salary/number of weeks in a month/ average number of hours worked per week = 195/4/48 = 1.01.

# HOW MUCH DO EGYPTIAN DATA WORKERS EARN? (2/3)

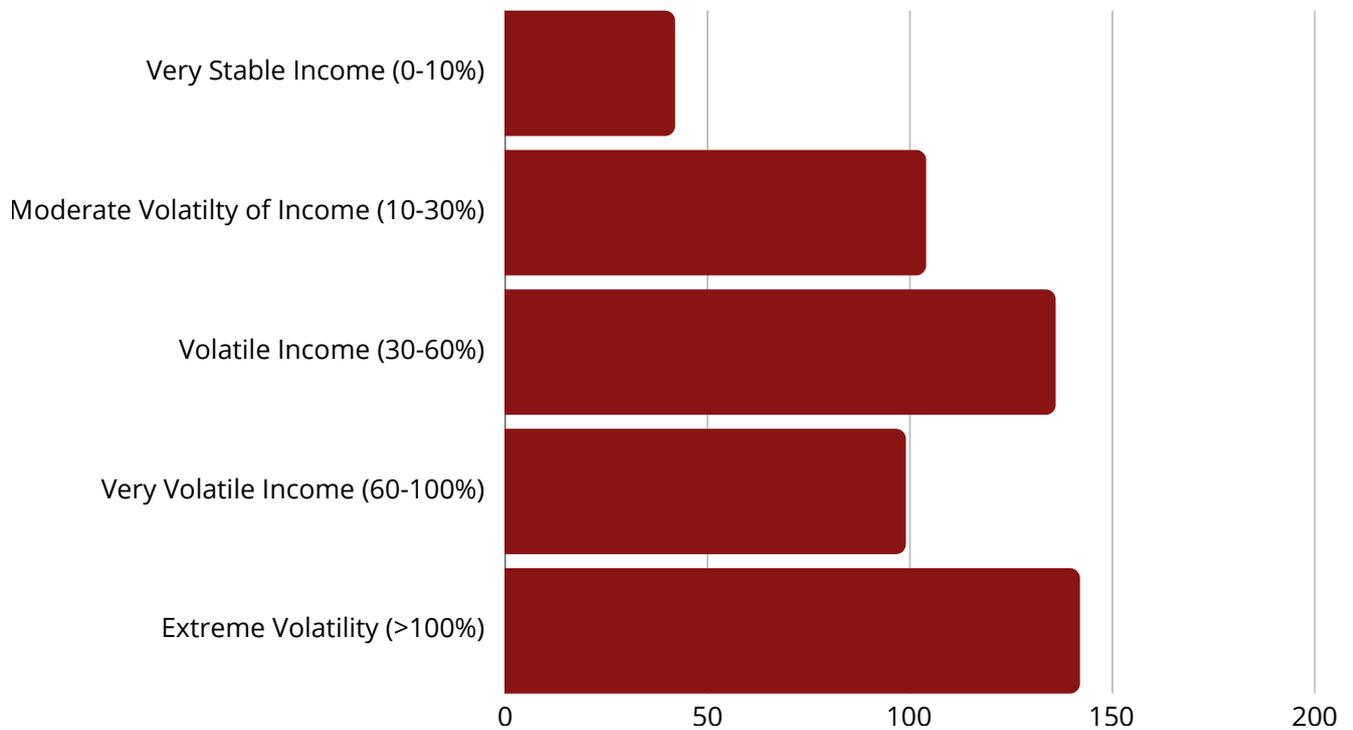

Figure 7 – Earnings Volatility (CV) *
*Assuming Missing Months = $0*

Figure 7 shows platform earnings volatility calculated over the last three months. The chart groups individuals into five brackets of income volatility (plus a sixth bracket for those without income), ranging from very stable earnings (0–10%) to extreme volatility (greater than 100%). As illustrated, **the number of people experiencing volatile or extremely volatile income is noticeably higher than those with very stable earnings**. Moderate and volatile income categories occupy the middle of the distribution, while extreme volatility stands out as the largest group, indicating that many individuals saw their earnings fluctuate dramatically over the period.

DiPLab

# HOW MUCH DO EGYPTIAN DATA WORKERS EARN? (3/3)

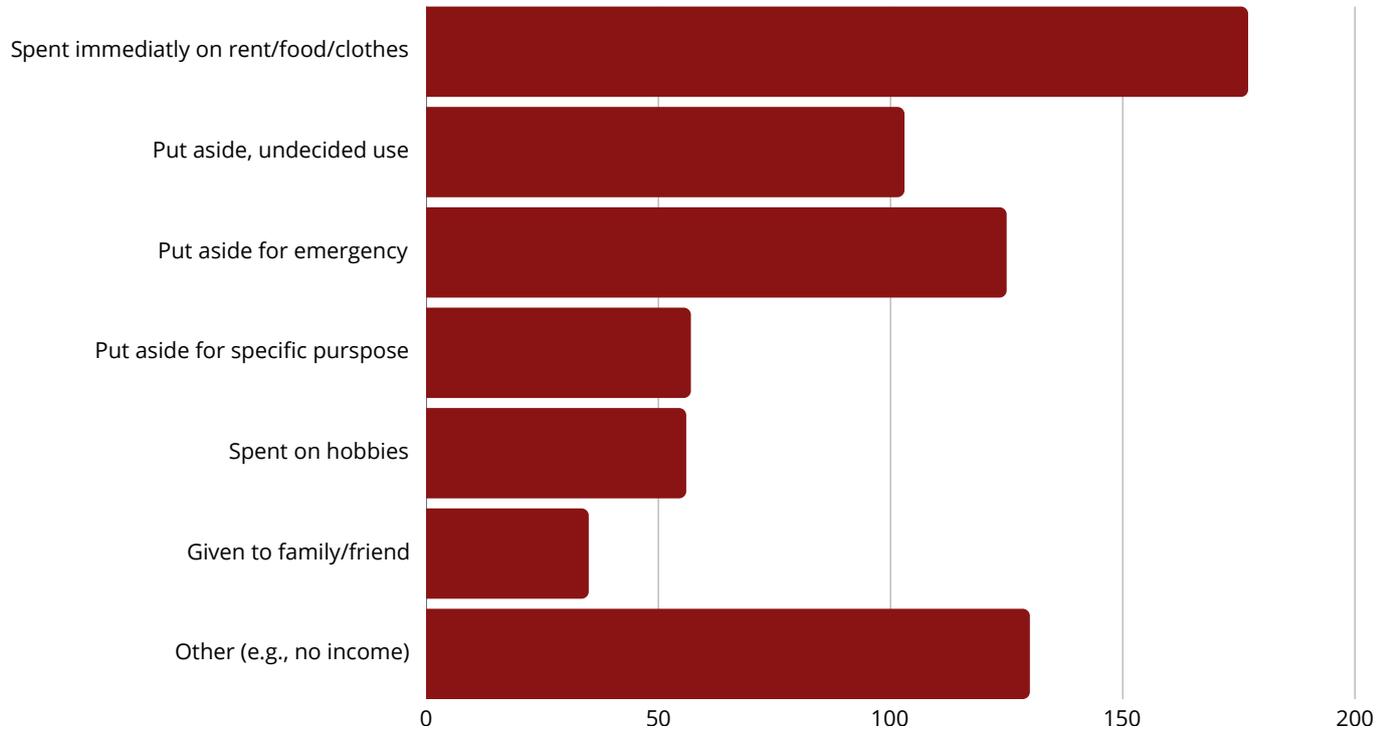

Figure 8 – How Workers Used Their Last Month's Earnings from Online Platforms

Figure 8 shows how **platform work primarily serves as a lifeline rather than a source of financial flexibility.** Only a small share of data workers used their earnings for discretionary purposes such as hobbies or holidays.

# GENDER (1/2)

Male workers make up the overwhelming majority of the Egyptian respondents. This may reflect various barriers to female participation in digital labor, concerns about online safety, inadequate legal protections, and cultural constraints. Despite this, no significant gender differences were observed in work patterns.

- On average, **women work on platforms 11.5 hours per week** (compared to 12.7 hours for men)
- **53% of women tend to work during the week**, 44% work during week-ends (which are roughly the same percentages as men).
- For Egyptian women, the **most common work hours on data work platforms are between 5 a.m. and 9 a.m.** which are the same as men.

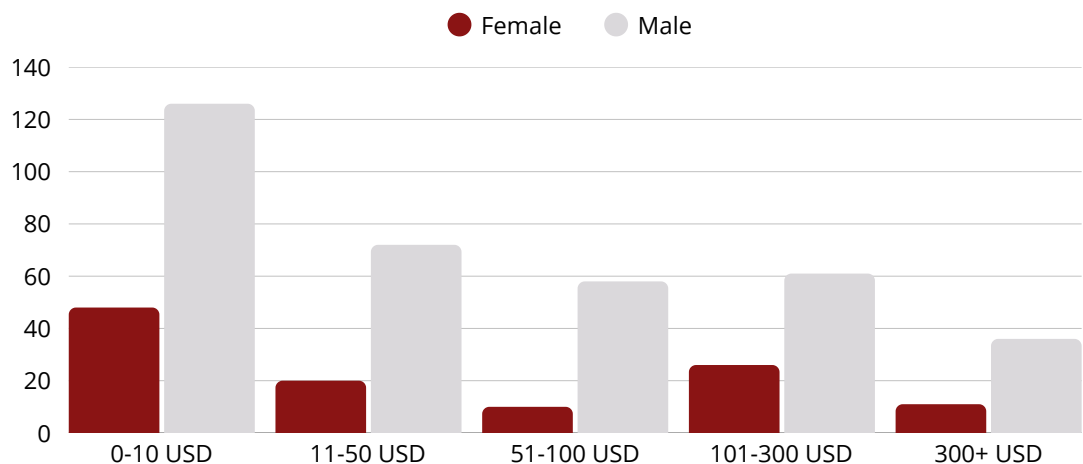

Figure 9 – Number of workers by earnings bracket by gender

Figure 9 chart shows that **lowest earning bracket (0–10 USD) is the largest among respondents, regardless of gender.** Although **some women appear in higher income brackets**, the overall distribution reflects a strong male dominance.

# GENDER (2/2)

The primary distinction appears to lie in the fact Egyptian women face greater challenges in maintaining their financial stability through platform work. The proportion of women without an additional job is twice that of men, and they tend to earn slightly less even as they exert more effort to look for new tasks everyday.

- **21% (n=31/152) of women have no other job**, compared to 10% (n=48/484) of men.
- In the three months prior to taking the survey, **women earned an average of USD 49** on platforms (N=152), while men earned USD 54 (N=484).
- **51% (n=77/152) of women check for new work more than once a day**, compared to 38% (n=182/484) of men

# MOTIVATING FACTORS FOR MICROWORK (1/2)

The **main motivation** that leads Egyptian workers to perform data work is the **need for money** (557 respondents). Another important motivator is a preference for working from home.

Platforms are seen as an **alternative income source due to the rise of precarious and informal work**. In addition to microtask platforms, an average of 15% of the participants have engaged in activities related to online sales, gambling, or sports betting, indicating that **data work is part of a larger ecosystem aimed at providing additional income on the internet**.

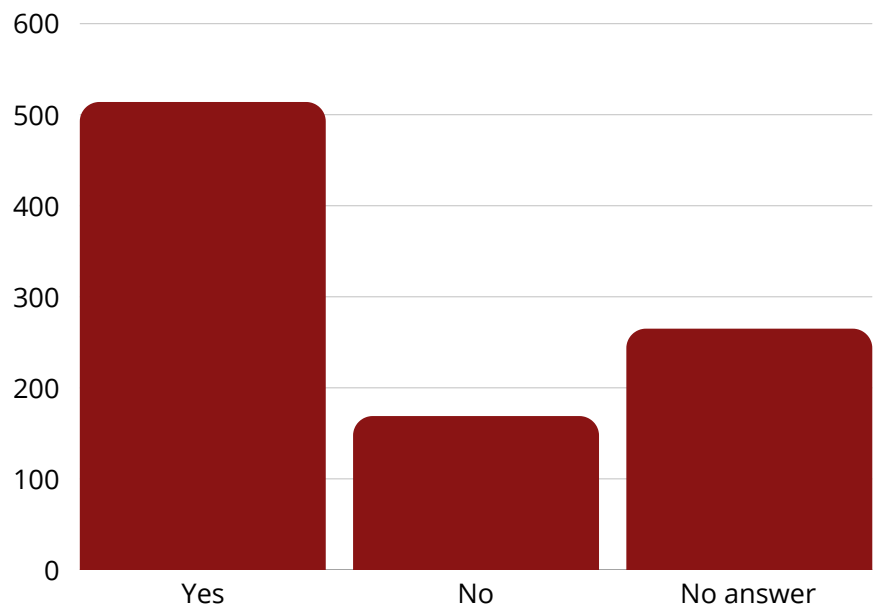

Figure 10 – Do data workers seek to secure a minimum monthly income on digital platforms to maintain a decent standard of living? (N=948)

# MOTIVATING FACTORS FOR MICROWORK (2/2)

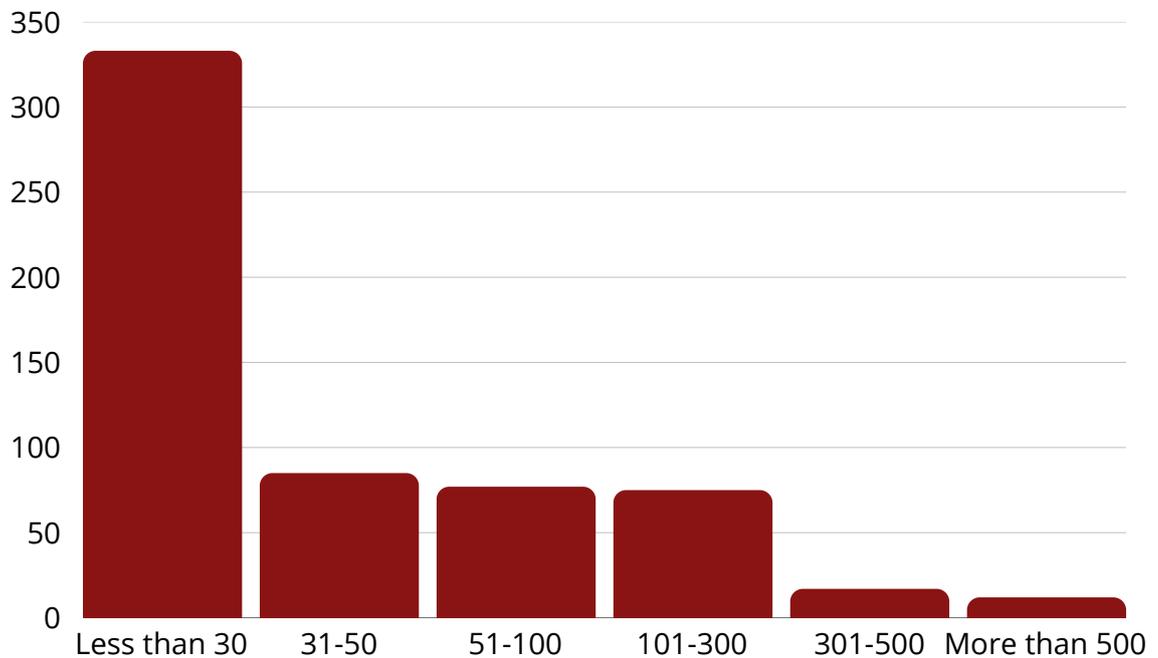

Figure 11 – Actual earnings from tasks carried out on digital platforms in the previous month (N=599)

Figures 10 and 11 **show that many respondents perceive their income as insufficient**. Figure 10 indicates that **more than half of respondents (54.2%) believe they need a minimum monthly income to live decently**, while 17.8% do not share this view and 28% did not respond. As displayed in Figure 11, **nearly half (48.75%) of respondents have earned less than USD 30 in the previous month**, with only a small fraction (4.18%) earning more than USD 300.

# DATA WORKERS' LINKS TO THE TECH SECTOR (1/2)

**Is data work related to a career in tech?** To answer this question we need to address the links between Egyptian platform workers and the tech sector at large: programming, engineering, internet marketing, digital design, digital strategy, research and development.

An overwhelming majority of respondents have ties to people working in the tech sector through intermediaries (usually friends), and most of **these ties are gender-specific. Essentially, they are man-to-man ties**, in that the respondent identifies as a man and the person he knows in the tech sector is a man as well.

The following figures show two main insights:
- Both informal platform work and technology jobs in Egypt fall generally in a **male-dominated field**. Woman-to-woman relations are a minority (7%), which is not surprising, but even relations across gender lines are less prevalent than man-to-man relations.
- **Connections in tech are overwhelmingly though friends**. This potentially shows that **technology work does not transcend generations** and tends to be important within the same demographic.

# DATA WORKERS' LINKS TO THE TECH SECTOR (2/2)

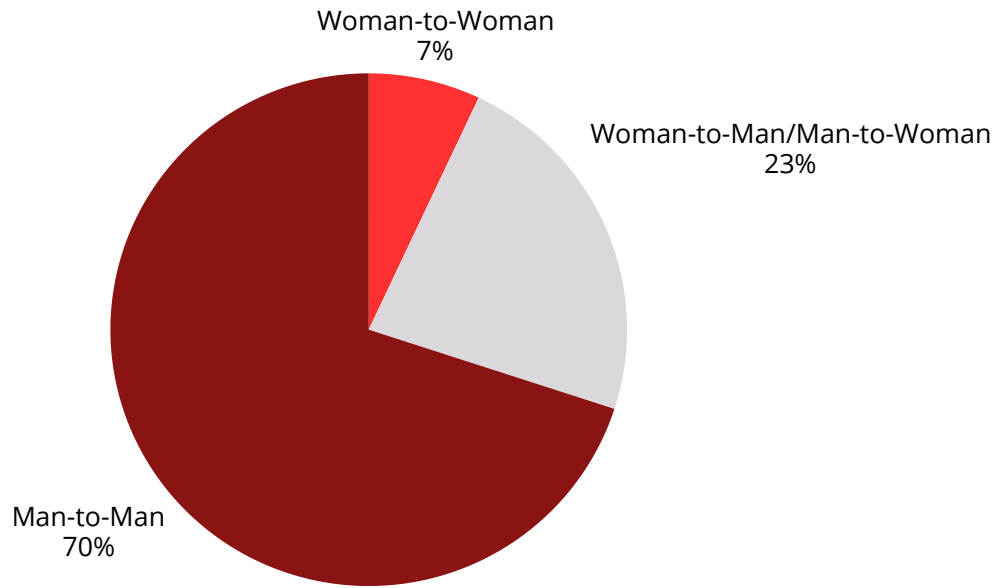

Figure 12 - Gender breakdown of tech connections

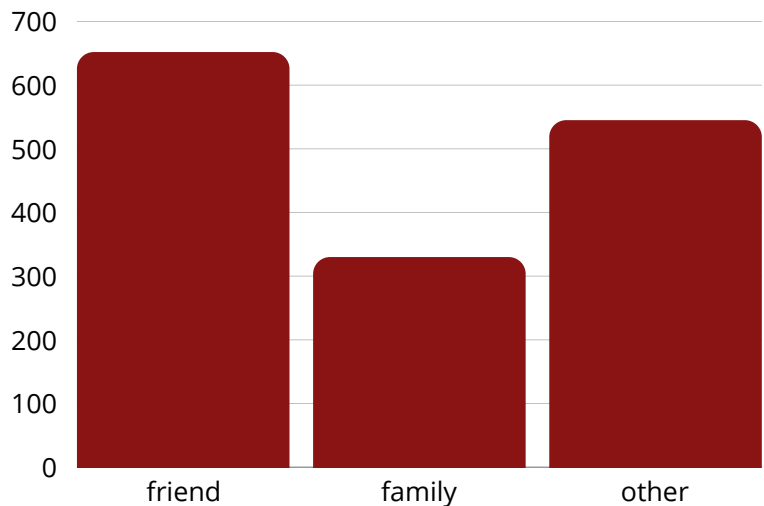

Figure 13 - Types of connections to the tech sector

# DATA WORKERS' MAIN COMPLAINTS: LOW WAGES

Preliminary findings indicate that platform payment systems create structural barriers to fair remuneration for workers. Data workers face three main types of obstacles:

1. **unreliable workflow and unsteady revenue** streams.
2. **opaque fee structures,** high commission fees, hidden costs, and unpredictable exchange rate fluctuations.
3. **delayed disbursements,** payment holds, withdrawal limits, and cross-border transfer barriers.

Other key findings include:

- **Earnings below livelihood thresholds**: A large share of workers earned less than $10 over the past three months.
- **Access inequality**: Less experienced workers are systematically disadvantaged by task allocation algorithms.

> "Payment is transferred from UHRS to Clickworker, and is kept in Clickworker for 39 days, then transferred to PayPal. Once funds are transferred to Paypal, I must wait until the beginning of the month for them to be transferred to my card." (Respondent F2MS3)

> "I had to create a card with the Egyptian post office (EasyPay); couldn't create a bank account because the banks I contacted told me they no longer deal with Paypal; EasyPay costs 120 EGP to create and fees are around 15 EGP every 3 months" (Respondent F2MS3)

# DATA WORKERS' MAIN COMPLAINTS: IDENTITY (1/2)

Egyptian data workers' **digital identities are shaped by algorithmic control and driven by economic demands**. These identities do not always align with who they are offline. Yet, within these constraints, **workers carve spaces for resistance, ethical agency, and autonomy**.

Data workers' concerns about identity cluster around three main themes:

1. **Professional and Personal Identity**: Egyptian data workers often feel like their professional identity is **reduced to metrics** (ratings, statistics, history if tasks accomplished) to which they tie their self-worth and they sometimes believe that hiding the degrees they have may favour their odds of getting work, which causes an identity conflict. Workers also mention ethical conflicts, they sometimes refuse tasks for ethical reasons.

> "Ethics are important. I refuse tasks that contradict my morals."
> (Respondent FG9)

> "Should I let go of my engineering degree and not mention it?"
> (Respondent FG11)

DiPLab

# DATA WORKERS' MAIN COMPLAINTS: IDENTITY (2/2)

2. **Socialization and Solidarity**: Data workers **perceive platforms as limiting solidarity** through bans, surveillance and segmentation. In response, they establish communities on other platforms, using national or cultural identities to enhance cohesion.

> "Platforms develop their tools to limit/ban more interactions between workers." (Respondent FG2)

> "Yes, I started UHRS group worldwide on FB – plus WhatsApp – now I have like 20 groups."(Respondent FG5)

3. **Algorithmic Visibility and Platform Influence**: Egyptian data workers perceive their nationality as negatively affecting both their visibility and pay rates. They also maintain a generally fraught relationship with platforms: they move easily between platforms without strong attachment to any single one, and they have observed that these services often penalize attempts at collective action, although such movements occasionally succeed.

> " Since I am Egyptian, I get tasks that have a very low pay rate (tasks that pay here 20$ pay overseas 450$!)."(Respondent FG9)

> "People complained collectively that the rate had decreased… and took a ban (regretful, painful incident)." (Respondent FG4)

# ARE EGYPTIAN DATA WORKERS BETTER OFF TODAY? (1/2)

The International Labor Organization (ILO) reported in 2018 that based on a survey of 3,500 data workers from 75 countries, including Egypt, the average age was 33.2 years. In developing countries, one in five workers were women. 37% of microworkers had a bachelor's degree and 20% held graduate degrees. Workers earned an average of $4.43 per hour and spent 18.6 hours per week on data work platforms. 32% of workers used these platforms as their primary source of income, and 36% worked seven days a week.

How do Egyptian data workers today compare with this historical data? Our findings reveal both **stagnation and significant shifts**: hourly rates remain frozen, and men still make up three-quarters of the workforce. However, the workforce has shifted from married adults with children to **highly educated single young people**, most of whom now hold part-time jobs but in occupations that are not considered high-skilled and use platforms for extra income. Despite this, three-quarters still rely on data work to pay their bills, earning less than half the minimum wage.

- In the ILO 2018 report, the average hourly rate for microtasks in developing countries was USD 4.43. According to our survey, today the Egyptian hourly rate amounts to $1,22[1].
- The ILO 2018 findings revealed that nearly **70% of users of the platform Microworkers identified as men globally**. According to our survey, today 76% of Egyptian data workers identify as men, indicating the prevalence of cultural barriers to women using microtasking platforms to earn their livelihoods.

---

[1] Average salary over the last 3 months /number of weeks per month/average number of hours worked per week = 58,76/4/12.3=1,22).

# ARE EGYPTIAN DATA WORKERS BETTER OFF TODAY? (2/2)

- Regarding marital status, 55% of data workers in developing countries were married, living with a partner or living in common-law marriage, and on average, 43% have children living in their households, according to ILO 2018 study. Our data display a stark change, probably due to the young age of the respondents and to their job instability: today in Egypt only 36% (n=222/610) of data workers are married, but 95% declare they live with minors in their household (although probably their siblings or members of extended families, not their own children; n=196/207).
- Furthermore, in 2018 57% of workers globally had completed **at least a bachelor's degree**, whereas in Egypt today, this percentage rises to 70% (n=496/671).
- In terms of healthcare, globally in 2018 **66% of data workers had some health insurance,** whereas today in Egypt, it is 59% (n=389/658).
- Among Egyptians with a formal job alongside data work, today **37% (n=137/370) work part-time**, slightly above the 2018 global average of 33%.
- Among Egyptians with a formal job in addition to data work, **only 12%** (n=81/666) **are currently engaged in what is usually pereceived by employers as 'high-skilled' occupations**[1]. This is significantly lower than the 2018 figures for Latin America and the Caribbean (65%), Asia and the Pacific (61%), and Europe (59%), but similar to the less than 20% observed in North America.

---

1 The categories were titled as follow: "I have no professional activity", "Employee", "Student", "Retired", "Manager and senior intellectual professional", "Craftsman, merchant or manager of a company", "Administrative employee" and intermediate occupation", "Worker", "Farmer". We consider "Managers and senior intellectual professional" as being a high-skilled occupation.

# EXPLORING EGYPTIAN DATA WORK: TRENDS AND IMPLICATIONS (1/2)

All the elements we have highlighted are closely tied to ongoing efforts to regulate technology that are taking place across continents. The attempts to regulate gig work in both Egypt and the Middle East have been welcomed by public opinion as necessary steps toward aligning technology with human values, social needs, and collective well-being.

However, these **regulatory endeavors have largely overlooked the vital roles played by data workers in AI development and their central position within the platform economy**. Numerous news stories and academic studies (including those by our team; Casilli et al. 2019, Viana et al. 2023, and Miceli et al. 2024) confirm that an increasing number of data workers are involved in producing data to train AI models. To keep costs low for this essential yet undervalued work, **tech companies often outsource it to low-income countries**, where labor is inexpensive and worker protections are minimal.

Microtasking platforms enable tech companies from the Global North to leverage the skills of workers to train their technological solutions. However, the benefits derived from this arrangement predominantly flow to wealthier nations, while countries like Egypt miss the opportunity to raise more tax income, and fail to provide extra welfare benefits to workers and to develop infrastructure.

Given that data work in Egypt is part of the **extensive global data supply chains, linking modest homes in Cairo to data centers in Northern California or Ireland**, it becomes clear that effective regulation of both informal platform labor and AI requires addressing and managing these complex global connections.

# EXPLORING EGYPTIAN DATA WORK: TRENDS AND IMPLICATIONS (2/2)

Our investigation into this form of work aims to illuminate the presence of data workers in Egypt and to assist local policymakers in gaining a deeper understanding of this emerging occupation within the intricate web of economic and political dependencies in which it is embedded.

The **Egyptian Labor Law 2025** (see Appendix 2) remains inadequate for data workers on digital platforms because it continues to rely on traditional employer–employee definitions, leaving platform workers outside the scope of social protection, minimum wage guarantees, and mechanisms for dispute resolution.

# GUIDELINES FOR POLICYMAKERS AND PLATFORMS (1/2)

The above challenges, specifically financial precarity, highlight policy needs and call for actionable reforms to improve transparency, access, and income security. Our recommendations are grouped into two categories: **regulatory reforms and proposed changes on the part of platforms**.

**GUIDELINES FOR PUBLIC POLICY**

**Improving the capacity of census and other official statistics to measure the prevalence of data work**, its evolution over time, and its place in the national labor market is a necessary step toward the development of evidence-based policies. To ensure cross-country comparability and mutual learning, it should be undertaken in collaboration with international agencies such as the International Labor Organization (ILO).

Promotion of **financial inclusion** would facilitate data workers' access to locally managed, affordable, and sustainable banking services. This would make earnings more predictable, reduce dependency on foreign and/or unofficial currency exchange providers, improve tax compliance, and ultimately contribute to bringing this activity into the formal economy.

**Simplifying activity registration processes** would also contribute to encourage formalization. To ensure that such a transition results in actual income improvements for data workers, complementary measures that ensure adequate coverage by social security would be particularly helpful.

# GUIDELINES FOR POLICYMAKERS AND PLATFORMS (2/2)

**GUIDELINES FOR PLATFORMS**

**Payments for platform-based data work should be more transparent**. Instead of letting clients set pay levels, accord bonuses, and validate (or reject) tasks in seemingly arbitrary ways, platforms should provide guidelines and advice that nudge decision-making in more worker-friendly directions. Beyond requiring payrates to be at or above minimum wage, platforms could set stricter limits to the possibility of rejecting tasks, insist on fair calculation of working time, and encourage payment of bonuses.

Platforms should also more transparently disclose the **criteria through which they allocate tasks across particular workers or groups of workers**. They should also encourage clients to offer their tasks to the largest pool possible, rather than facilitating selection along often strict criteria.

**Platforms should put in place conflict resolution mechanisms based on fairness principles.** Data workers' complaints and queries should have greater priority than is common, possibly through a mediation system in case of disagreements or disputes with clients. Also, dedicated support staff should be available to help workers who experience technical issues.

*These recommendations offer only some high-level policy directions. We hope they open space for ongoing dialogue with stakeholders and for future insights to further support data workers and enhance their working conditions.*

DiPLab

# APPENDIX 1: METHODOLOGY (1/5)

Using a mixed methods approach, this study combines a quantitative survey of over 600 Egyptian data workers with qualitative insights from focus groups. The study also integrates ethnographic insights through the researchers' immersion in online social media groups frequented by Egyptian data workers. This design enables a comprehensive analysis of several aspects including:

- **Survey Data**: The survey data presented was collected via LimeSurvey and administered in Arabic to populations of Egyptian platform workers working on two distinct microtasking platforms (Microworkers and Clickworker). A total of **639 completed responses** were obtained (and an additional 308 partial responses), with **76% identifying as male and 24% as female**. The questionnaire comprised 12 thematic sections covering topics such as digital platform use, work patterns, income and payment practices, digital literacy, household composition, and sociodemographics. It included **133 questions**, with an average completion time of 30 minutes. Participants were recruited through tasks published on both platforms. The survey received the approval of the university of Angers's Research Ethics Board.

| Platform | Number of Responses | Dates |
|---|---|---|
| Microworkers | 347 complete<br>143 partial | May 13 to September 10, 2024. |
| Clickworker | 292 complete<br>166 partial | August 16 to August 31, 2024 |

DiPLab

# APPENDIX 1: METHODOLOGY (2/5)

| Topic | Number of questions |
|---|---|
| Digital Platform Awareness, Usage History, and Digital Income-Generation Activities | 44 |
| Digital Device Ownership, Accessibility Tools, and Computer Literacy Assessment | 3 |
| Internet Access Patterns and Frequently Visited Websites | 2 |
| Work Patterns, Task Selection Behavior, and Community Engagement in Digital Labor | 22 |
| Income Patterns, Financial Requirements, and Payment Management in Digital Labor | 6 |
| Educational Qualifications and Language Proficiency Assessment | 6 |
| Employment Status, Professional History, and Work Arrangements | 19 |
| Working Conditions and Job Quality Assessment: Current and Previous Employment | 6 |
| Household Composition and Domestic Responsibilities | 5 |
| Personal and household income composition, financial challenges | 10 |
| Position generator | 1 |
| Sociodemographics | 9 |

DiPLab

# APPENDIX 1: METHODOLOGY (3/5)

- **Focus Groups**: To complement the survey data, a series of focus groups were conducted in 2025 with Egyptian data workers to explore in greater depth their experiences with platform-based payment systems and perception of their professional identity. In total, 6 focus groups were organized around the theme of professional identity (totaling **15 participants, 9 men and 6 women**) and 4 organized around the theme of payments (totaling **11 participants, 8 men and 3 women)** participated in these focus groups. Each focus group lasted on average 90 minutes. Focus groups participants were recruited via social media campaigns and were conducted online in Arabic language using Microsoft Teams. The focus groups were recorded and then transcribed.

- Focus groups revolved around two topics: professional identity and worker compensation. The focus group around professional identity was organized in three themes : **(1) Personal & Professional Identity** [8 questions] **(2) Social Capital, Collective Identity & Solidarity** [11 questions] **(3) Algorithmic Visibility & Platform Influence** [6 questions]. The focus groups around compensation was organized in four themes : **(1) Platform and payment methods**, **(2) Earnings and payment structures**, **(3) Challenges in receiving payments**, and **(4) Perspectives on improvements and worker status**.

DiPLab

# APPENDIX 1: METHODOLOGY (4/5)

Focus groups - TOPIC: PROFESSIONAL IDENTITY

| FG | Date | Number of Participants | Codes of participants |
|---|---|---|---|
| 1 | Mar 22, 2025 | 2 | FG1, FG2 |
| 2 | Mar 25, 2025 | 3 | FG3, FG4, FG5 |
| 3 | Mar 28, 2025 | 3 | FG6, FG7, FG8 |
| 4 | Aug 19, 2025 | 1 | FG0 |
| 5 | Aug 21, 2025 | 4 | FG10, FG11, FG12, FG13 |
| 6 | Aug 23, 2024 | 2 | FG14, FG15 |

Focus groups - TOPIC: COMPENSATION

| FG | Date | Number of Participants | Codes of participants |
|---|---|---|---|
| 1 | Jan 15, 2025 | 3 | F1MS2, F1MS3, F1FS4 |
| 2 | Jan 15, 2025 | 2 | F2MS3, F2MS4 |
| 3 | Jan 31, 2025 | 3 | F3MS5, F3MS6, F3FS4 |
| 4 | Feb 2, 2025 | 2 | F4FS5, F4MS7, F4MS8 |

DiPLab

# APPENDIX 1: METHODOLOGY (5/5)

**Online forum monitoring**: Researchers practiced online monitoring of social media groups frequented by Egyptian data workers (WhatsApp groups, Facebook, Telegram and Messenger communities). This offered **contextual understanding** of workers' lived experiences, informal exchanges, and adaptive strategies beyond formal surveys and interviews.

*Data collection has been conducted in compliance with GDPR provisions. Moreover, the guidelines for research integrity of the Université d'Angers (France) have informed our research protocol.*

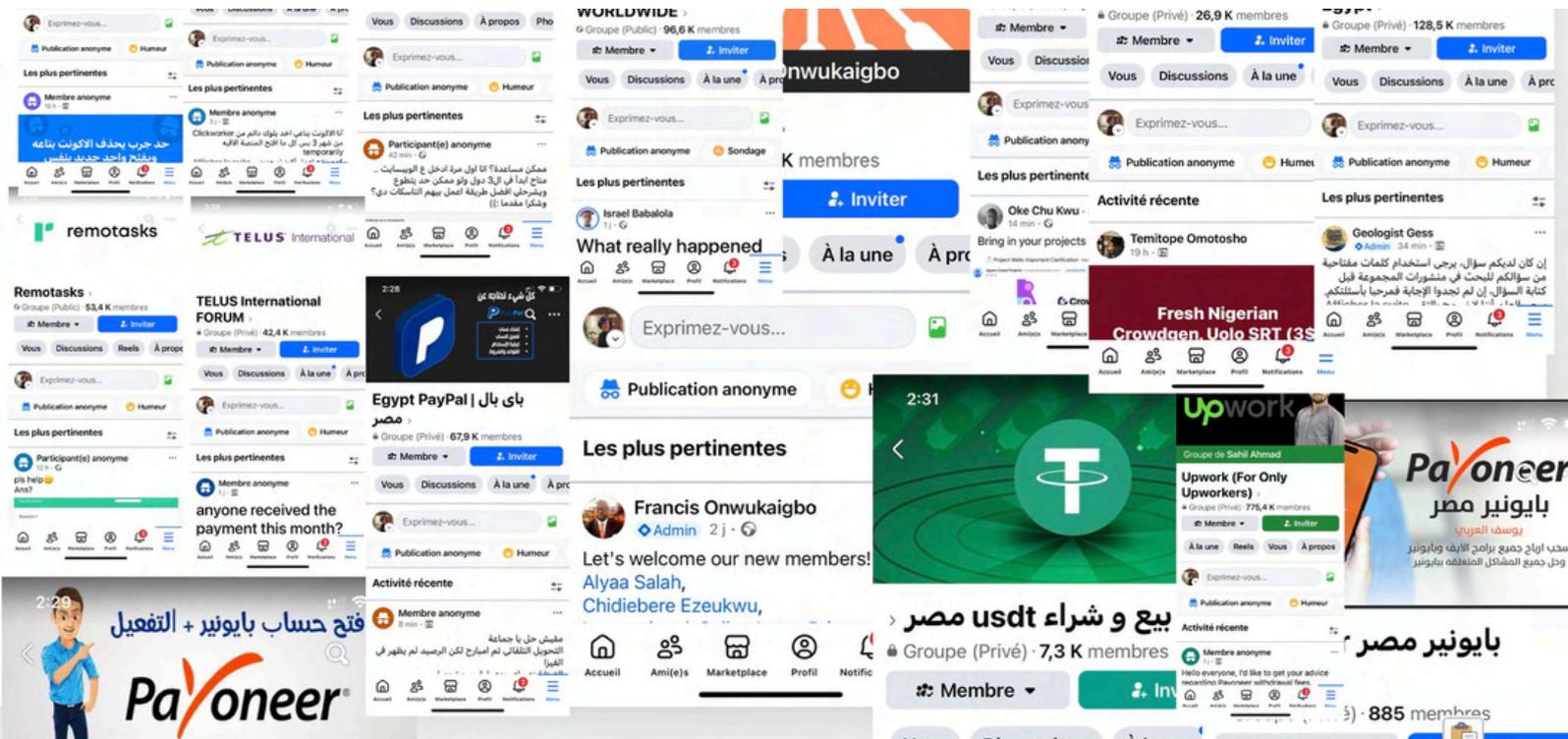

# APPENDIX 2: KEY HIGHLIGHTS FROM 2025 EGYPTIAN LABOR LAW (قانون العمل رقم 14 لسنة 2025)

1. No explicit recognition of platform work or gig workers

2. Digital labour is implicitly absorbed under "non-regular" or "informal" work

3. Creation of a new national fund for non-regular workers

4. No provisions for algorithmic management or data governance

5. No rules on platform payment practices

6. Employment classification remains ambiguous

7. Explicit provisions for "new forms of work" do not mention digital platforms

8. Foreign digital workers remain unregulated

While the law expands rules for foreign workers, it does not address:
• Egyptians working for foreign platforms
• Foreigners working remotely from Egypt

9. Social protection improvements do not include platform obligations

DiPLab

# TEAM

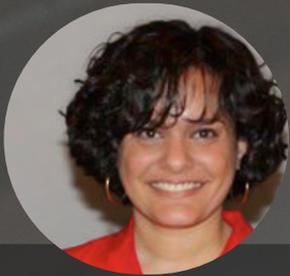

**Myriam Raymond**
Researcher at GRANEM (Groupe de Recherche ANgevin en Économie et Management). Associate researcher at LEMNA (Digital and Organisational Transformation), and associate researcher at the research group DiPLab (Digital Platform Labor). E-mail: myriam.raymond@univ-angers.fr

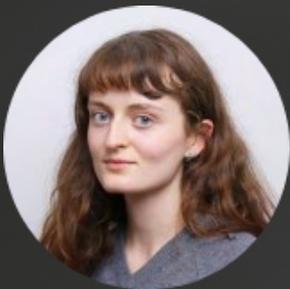

**Lucy Neveux**
Dual-degree student at HEC Paris, France's leading business school, and ENSAE Paris, specializing in statistics, economics, and data science. E-mail: lucy.neveux@hec.edu

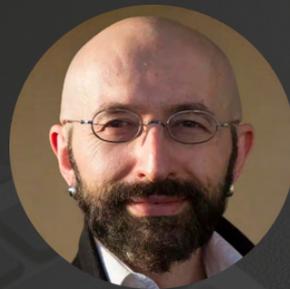

**Antonio A. Casilli**
Professor of Sociology at the Polytechnic Institute of Paris and affiliate at the Minderoo Centre for Technology & Democracy, University of Cambridge. Co-founder of DiPLab (Digital Platform Labor) and of the International Network on Digital Labor (INDL). E-mail: antonio.casilli@ip-paris.fr

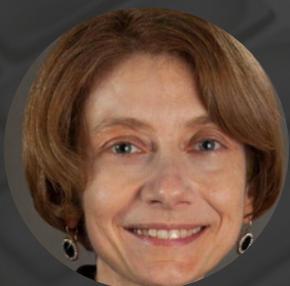

**Paola Tubaro**
Research professor (Directrice de Recherche) in sociology at the National Centre for Scientific Research (CNRS) and member of the Center for Research in Economics and Statistics (CREST) in France. She is the co-founder of DiPLab. E-mail: paola.tubaro@cnrs.fr